\documentclass[10pt]{article}
\usepackage{graphicx}
  \usepackage{amsmath,amsthm, amssymb, latexsym}
  \usepackage{overpic}
  \usepackage{xcolor}
  \usepackage{hyperref}

\usepackage{tikz}
\usetikzlibrary{patterns}
  \usetikzlibrary{snakes}
 \usetikzlibrary{arrows}
 \usetikzlibrary{decorations.markings}
 \tikzset{
    photon/.style={decorate, decoration={snake}},
    electron/.style={ postaction={decorate},
        decoration={markings,mark=at position .55 with {\arrow[]{>}}}},
    gluon/.style={decorate,
        decoration={coil,amplitude=3pt, segment length=6pt}} 
}

\def\Title#1{\begin{center} {\Large #1 } \end{center}}
\def\Author#1{\begin{center}{ \sc #1} \end{center}}
\def\Address#1{\begin{center}{ \it #1} \end{center}}

\newcommand\pubblock{\rightline{\begin{tabular}{l} Proceedings of the Second Annual LHCP\\ \pubnumber\\
         \pubdate  \end{tabular}}}

\newenvironment{Abstract}{\begin{quotation} \begin{center} 
             \large ABSTRACT \end{center}\bigskip 
      \begin{center}\begin{large}}{\end{large}\end{center} \end{quotation}}

\newenvironment{Presented}{\begin{quotation} \begin{center} 
             PRESENTED AT\end{center}\bigskip 
      \begin{center}\begin{large}}{\end{large}\end{center} \end{quotation}}

\def\Acknowledgements{\bigskip  \bigskip \begin{center} \begin{large}
             \bf ACKNOWLEDGEMENTS \end{large}\end{center}}

\def\beq{\begin{equation}}
\def\eeq#1{\label{#1}\end{equation}}
\def\eeqn{\end{equation}}

\def\beqa{\begin{eqnarray}}
\def\eeqa#1{\label{#1}\end{eqnarray}}
\def\eeqan{\end{eqnarray}}

\let\bar=\overbar

\def\Dslash{\not{\hbox{\kern-4pt $D$}}}
\def\dslash{\not{\hbox{\kern-2pt $\del$}}}

\def\msb{{\bar{\ssstyle M \kern -1pt S}}}

\usepackage{color}

\newcommand\pubnumber{ ATL-PHYS-PROC-2014-103 }

\newcommand\pubdate{\today}

\def\affiliation{
On behalf of the ATLAS Collaboration, \\
SLAC, Stanford University  }

\begin{document}

\large
\begin{titlepage}
\pubblock

\vfill
\Title{  Jet Charge\\ {\normalsize with the ATLAS Detector using $\sqrt{s}=8$ TeV $pp$ Collision Data } }
\vfill

\Author{ Benjamin Nachman}
\Address{\affiliation}
\vfill
\begin{Abstract}

The momentum-weighted sum of the charges of tracks associated to a jet provides an experimental handle on the electric charge of fundamental strongly-interacting particles.  An overview of a study of this \textit{jet charge} observable for jets produced in dijet and semileptonic $t\bar{t}$ events using $5.8$ $\mathrm{fb}^{-1}$ of data with the ATLAS detector at $\sqrt{s}=8$ TeV is described here.  In addition to providing a constraint on hadronization models,  jet charge has many possible applications in measurements and searches.  The modelling of jet charge and its performance as a charge-tagger are studied in order to establish this observable as a tool for future physics analyses.  

\end{Abstract}
\vfill

\begin{Presented}
The Second Annual Conference\\
 on Large Hadron Collider Physics \\
Columbia University, New York, U.S.A \\ 
June 2-7, 2014
\end{Presented}
\vfill
\end{titlepage}
\def\thefootnote{\fnsymbol{footnote}}
\setcounter{footnote}{0}
%

\normalsize 


\section{Introduction}

Due the confining nature of the strong force, a quark's electric charge is not directly observable.  Only color singlet hadrons can be measured as schematically illustrated in Fig.~1.  However, the momentum-weighted sum of the charges of tracks associated to a jet provides an experimental handle on the electric charge of fundamental, strongly-interacting particles.  This {\it jet charge} observable was first proposed by Field and Feynman~\cite{Feynman1978} and has been used in many experimental studies, most recently in the measurement of the top quark charge at the Tevatron~\cite{Abazov2007,CDF2011} and the Large Hadron Collider (LHC)~\cite{ATLAS2011,CMS2012}.  Recent phenomenological studies have suggested further uses of the jet charge at the LHC, including tagging the charge of heavy particles like $W$ bosons or as of yet undiscovered $W'/Z'$ bosons~\cite{Krohn2012}.  Furthermore, there are new methods for calculating the scale evolution of the jet charge distribution, in analogy to the evolution for parton distribution functions (PDF)~\cite{Waalewijn2012,Krohn2012}. Thus, a measurement of jet charge could provide new insight into the perturbative nature of Quantum Chromodynamics (QCD).  An overview of an experimental study of the modelling and performance of jet charge as a charge-tagger in hadronic final states with the ATLAS detector~\cite{atlas} at the LHC is presented here (see Ref.~\cite{jetcharge} for additional details).

\vspace{-2mm}

\begin{figure}[h!]
\begin{overpic}[scale=0.5]{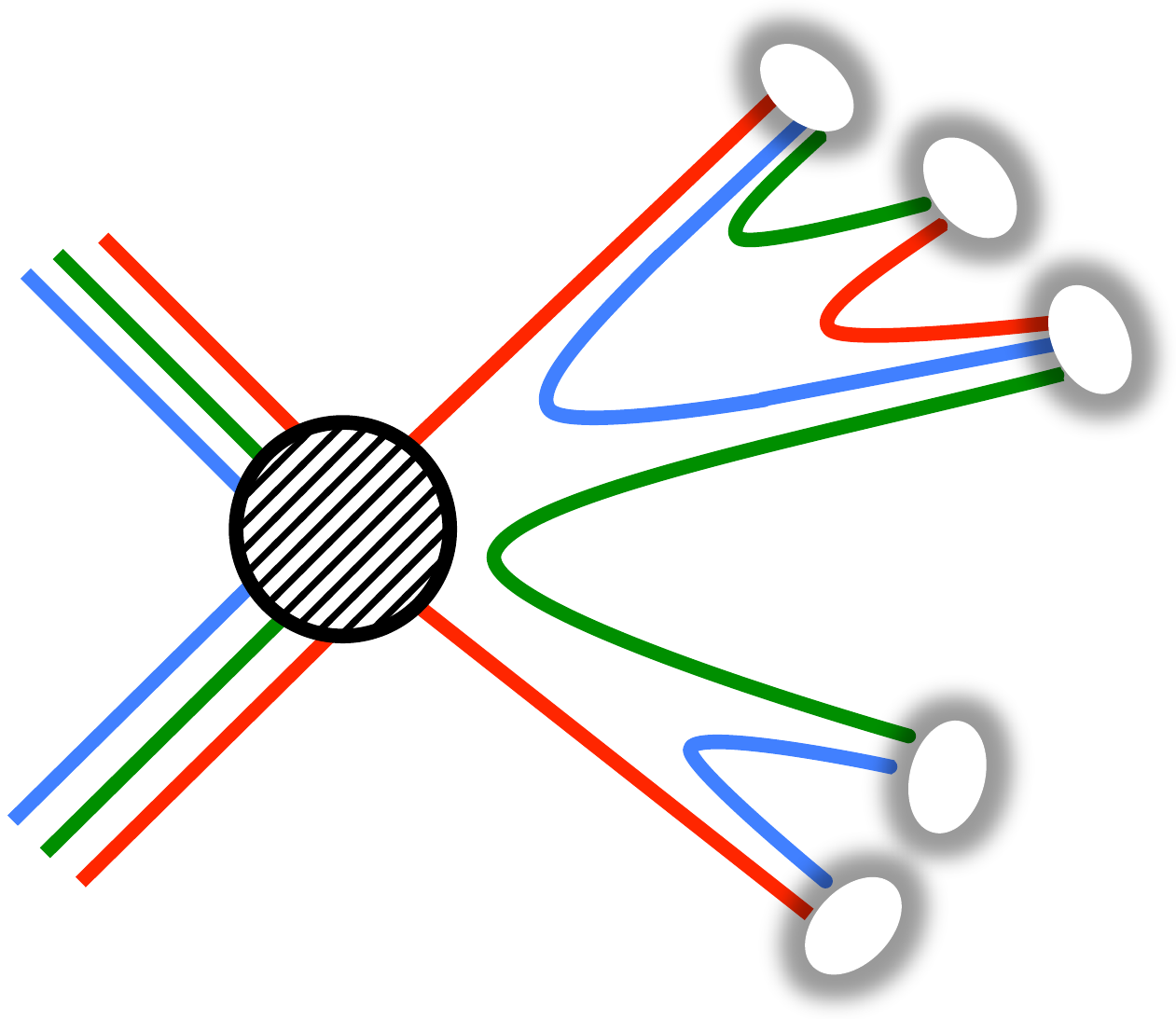}
\put(31,64){\large $u$}
\put(33,23){\large $\bar{d}$}
\put(28,57){$2/3e$}
\put(29,16){$1/3e$}
\put(67,77){$p$}
\put(80,68){$\pi^+$}
\put(91,55){$\bar{p}$}
\put(78,18){$\pi^-$}
\put(70,6){$\pi^+$}
\put(130,40){$\text{\Large$ \color{black}Q = \frac{1}{({p_T}_j)^\kappa}{\color{black}\sum_{i\in \text{\bf Tr}} (p_T^i)^\kappa} \times {\color{black} q_i}$  \hspace{5mm} (1)}$}
\end{overpic}
\caption{ Schematic diagram showing the unobservable quark charge transmitted to the charge of hadrons.}
\end{figure}

\vspace{-1mm}

There is no unique way for defining the jet charge.  In the studies presented here, the definition for the jet charge $Q$ is given in Eq.~1, where the sum runs over all tracks ghost-associated~\cite{area} to a particular jet $j$.  The charges of the tracks $q_i$ are regulated by the factor $(p_T^i/{p_T}_j)^\kappa$ with a tuning parameter $\kappa$.  As $\kappa\rightarrow 0$, the jet charge is highly sensitive to soft tracks which are not measured, and as $\kappa\rightarrow\infty$, the jet charge is only sensitive to the charge of the leading track.  For this study, jets are constructed from topological calorimeter clusters~\cite{TopoClusters} and clustered using the anti-$k_t$ algorithm with radius parameter $R=0.4$~\cite{Cacciari:2008gp}.   The tracks used to build the jet charge are required to have $p_T>500$ MeV in addition to a set of quality criteria rejecting tracks from pileup and cosmic rays.  For further details about the object selection, please see Ref.~\cite{jetcharge}.

Two physics processes are used to study the charge tagging performance of jet charge: $t\bar{t}$ and QCD dijets.  Events with the pair production of top quarks and one lepton in the final state are ideal for a data-driven measurement of the jet charge performance.  In particular, the charge of the hadronically decaying $W$ boson should be opposite to the charge of the measured lepton.  Such events are selected using standard techniques ($b$-tagging, missing transverse momentum, and transverse mass) and are required to have a pair of jets with dijet invariant mass within 30 GeV of the $W$ boson mass.  These jets are labeled as $W$ daughters.  Top quark pair production is simulated using the next-to-leading order (NLO) generator {\tt MC@NLO}~4.06 \cite{Frixione:2002ik,Frixione:2003ei} with the NLO PDF set {\tt CT10}~\cite{Lai:2010vv,Gao:2013xoa}, and parton showering and underlying event modelled with fortran {\tt HERWIG}~\cite{Corcella:2000bw} and {\tt JIMMY}~\cite{JIMMY}, respectively.  For further details about the selection and MC modelling, please see Ref.~\cite{jetcharge}.

Events with QCD dijet production are used both for MC-based performance studies and to qualitatively probe the modeling of the jet charge distribution.  The {\it flavor} of a jet is defined as the particle type of the highest energy parton in the MC record within the jet cone.   QCD dijet events are identified using a variety of single jet triggers and are modelled using {\tt PYTHIA} 8~\cite{Sjostrand2008852}.  For both the QCD and the $t\bar{t}$ selections, the dataset used corresponds to $5.8$ fb${}^{-1}$ collected at $\sqrt{s}=8$ TeV in 2012 with the ATLAS detector.

\section{Results}

Figure 2 shows the sum of the jet charges of both $W$ daughters from the semi-leptonic $t\bar{t}$ selection.  The dijet charge is plotted separately for $\mu^+$ (dijet charge shifted negative) and $\mu^-$ (dijet charge shifted positive).  The MC predicts well the shape of the jet charge distribution (MC normalized to the data) within the uncertainties, including jet energy scale and resolution uncertainties as well as an uncertainty in the tracking efficiency.  The separation between positive and negative hadronically decaying $W$ bosons is quantified in the right plot of Fig. 2 - a negative $W$ rejection (inverse of efficiency) of about 5 is expected for a positive $W$ boson efficiency of 50\%.  Figure 3 shows that the separation between charges is maintained (in simulation) even at large $W$ $p_T$ where large radius jets are used as the $W$ daughters become merged.  

\begin{figure}[h!]
\begin{center}
\includegraphics[width=0.45\textwidth]{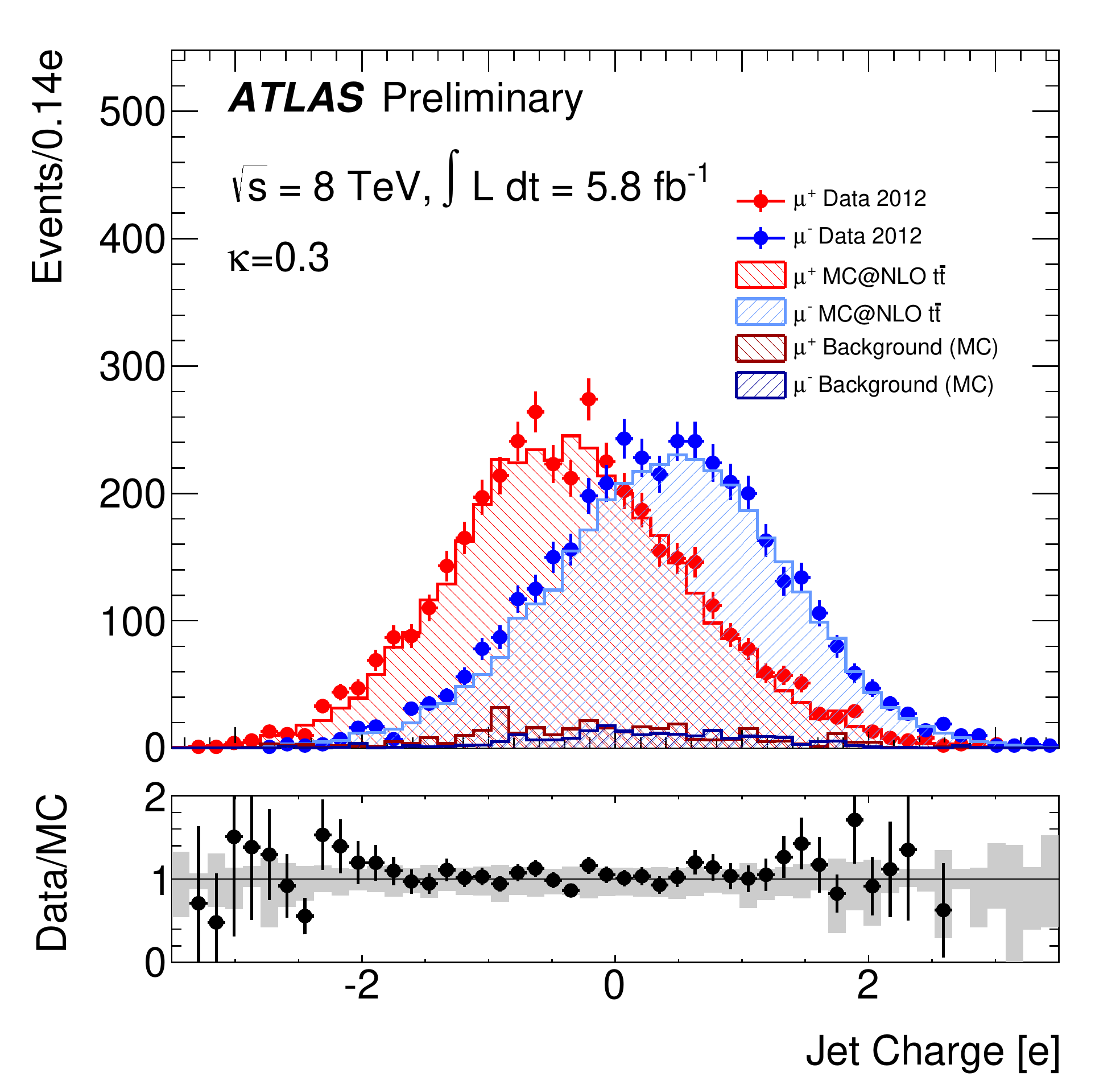}\includegraphics[width=0.45\textwidth]{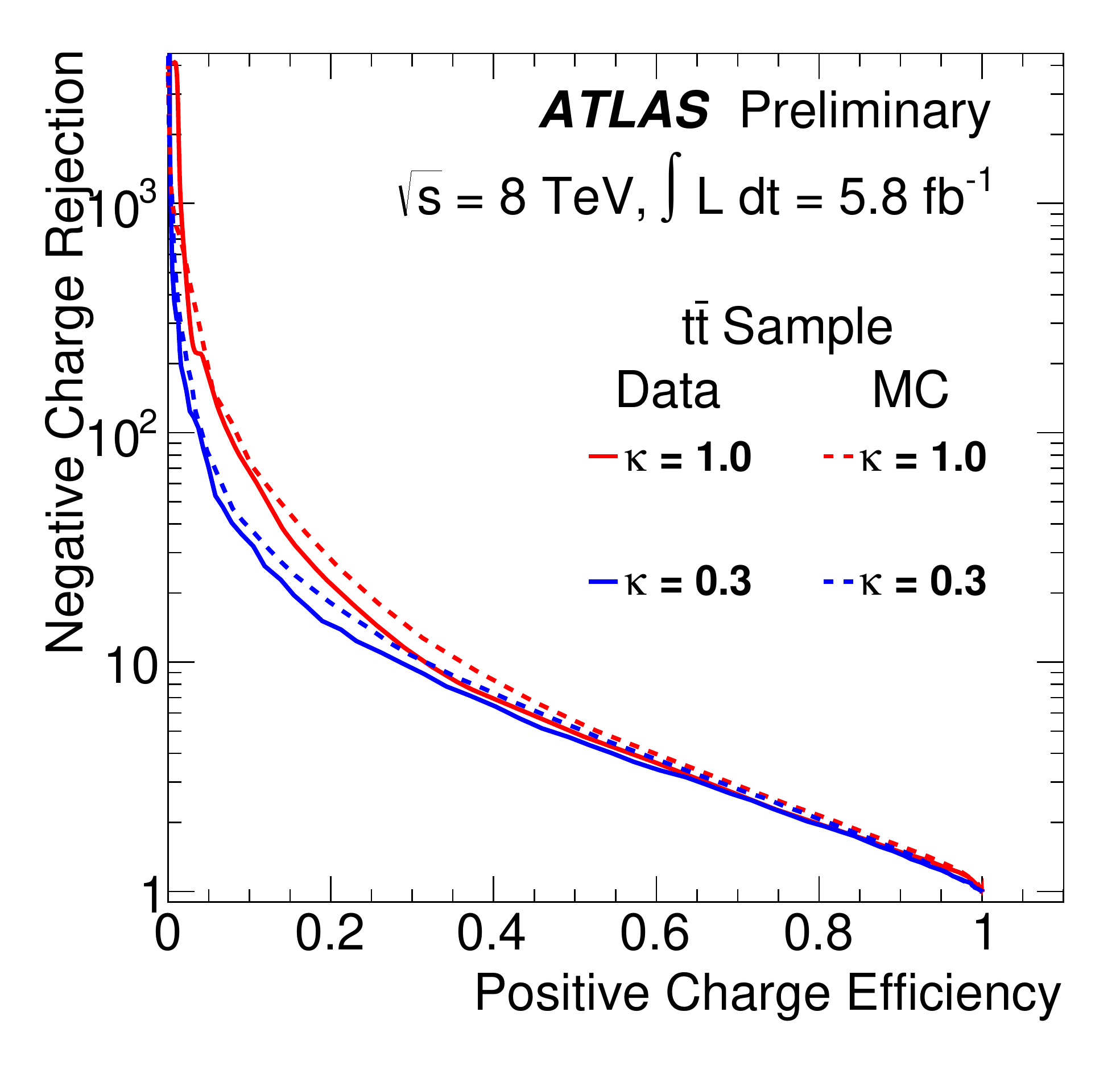}
\end{center}

\vspace{-5mm}

\caption{The sum of $W$ daughter jet charges (left) and the quantified charge tagging performance (right).}
\end{figure}

\begin{figure}[h!]
\begin{center}
\includegraphics[width=0.45\textwidth]{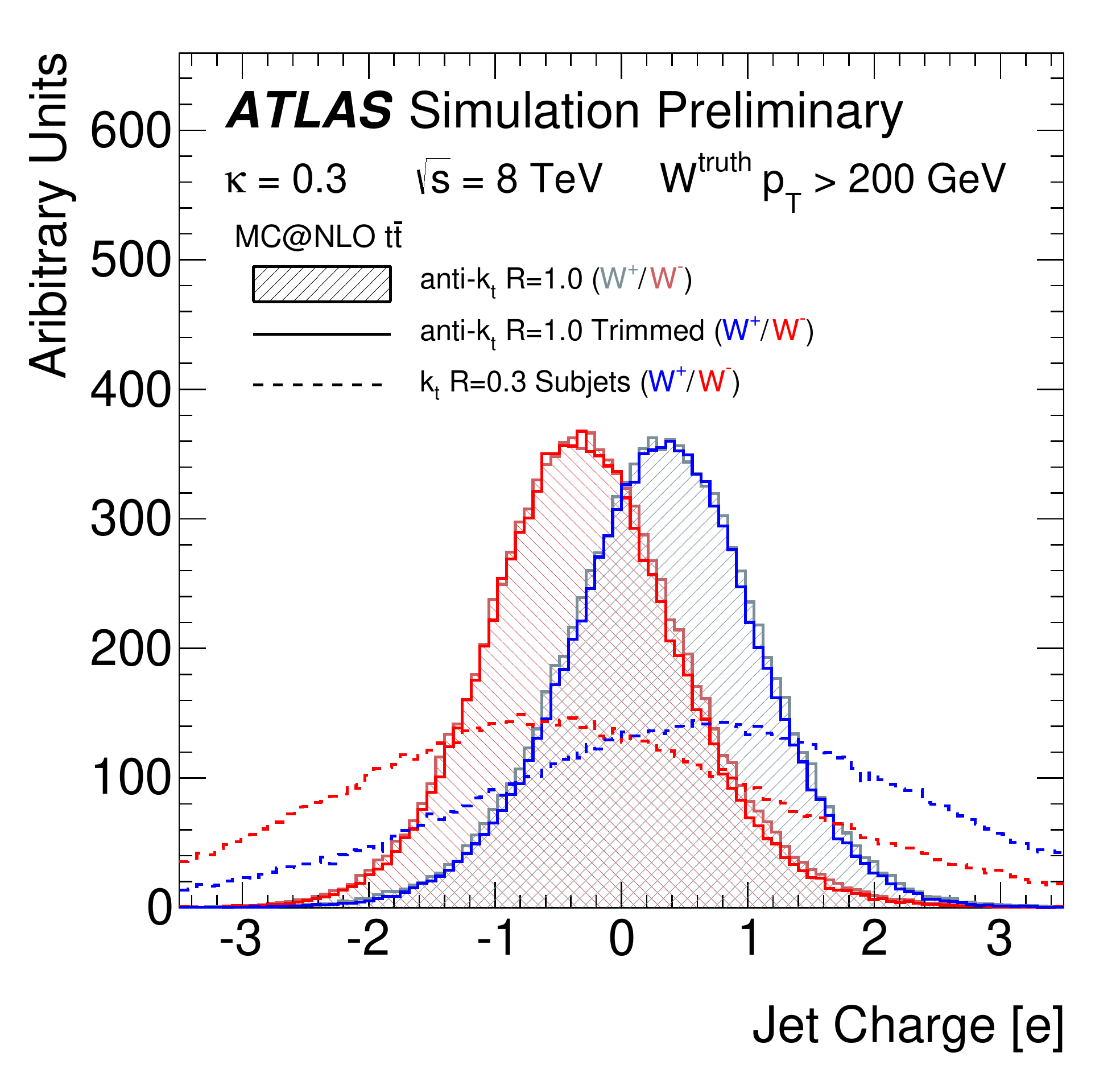}
\end{center}

\vspace{-5mm}

\caption{The jet charge distribution in $t\bar{t}$ at high $W$ $p_T$, constructed from tracks associated to the $R=1.0$ jet.  There is only a small impact from trimming, since tracks are well associated to the primary vertex.  The large $R$ jet charge performs better than the sum of the subjet charges.}
\end{figure}

\newpage

The jet charge distribution for various jet flavors is shown in Fig. 4.  There is some discrimination between quark types of opposite charge, but there is not much distinction between quark types of the same sign of charge. Jet charge may be useful for quark/gluon tagging since it is well modeled, but Fig. 4 shows that the separation between quarks and gluon is not large.  The charge tagging performance for distinguishing the charge of a jet's flavor is quantified in Fig. 5.  The plots in Fig. 5 indicate that the charge tagging efficacy and optimal $\kappa$ value do not have a strong $p_T$ dependance.

\begin{figure}[h!]
\begin{center}
\includegraphics[width=0.5\textwidth]{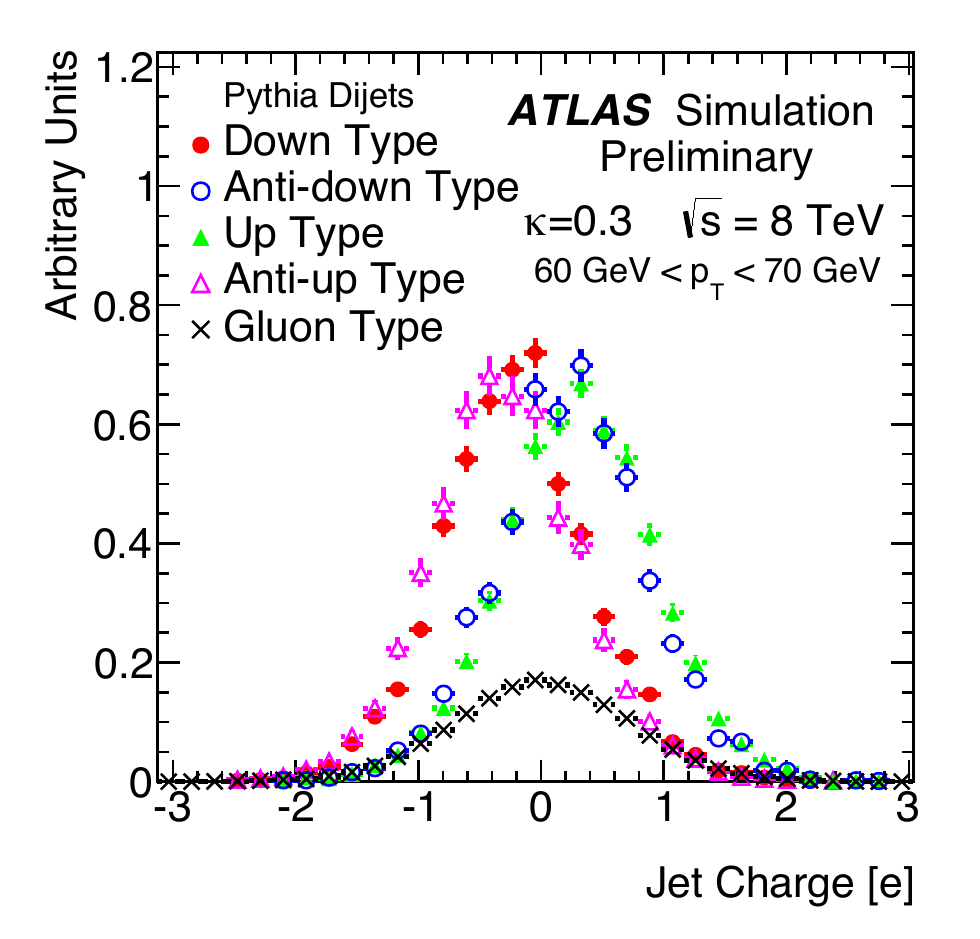}
\end{center}

\vspace{-7mm}

\caption{The jet charge distribution for various jet flavors.  Note that the gluon distribution is scaled down while the quark type distributions have the same normalization.}
\end{figure}

\begin{figure}[h!]
\begin{center}
\includegraphics[width=0.5\textwidth]{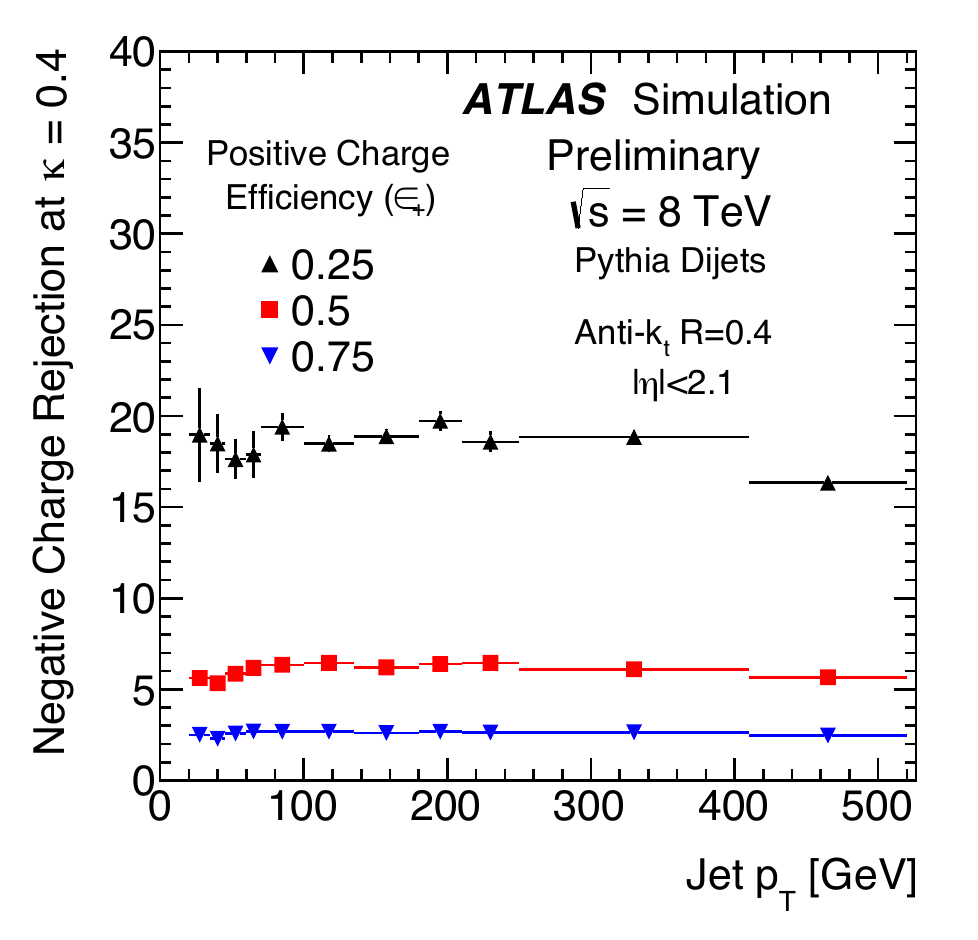}\includegraphics[width=0.5\textwidth]{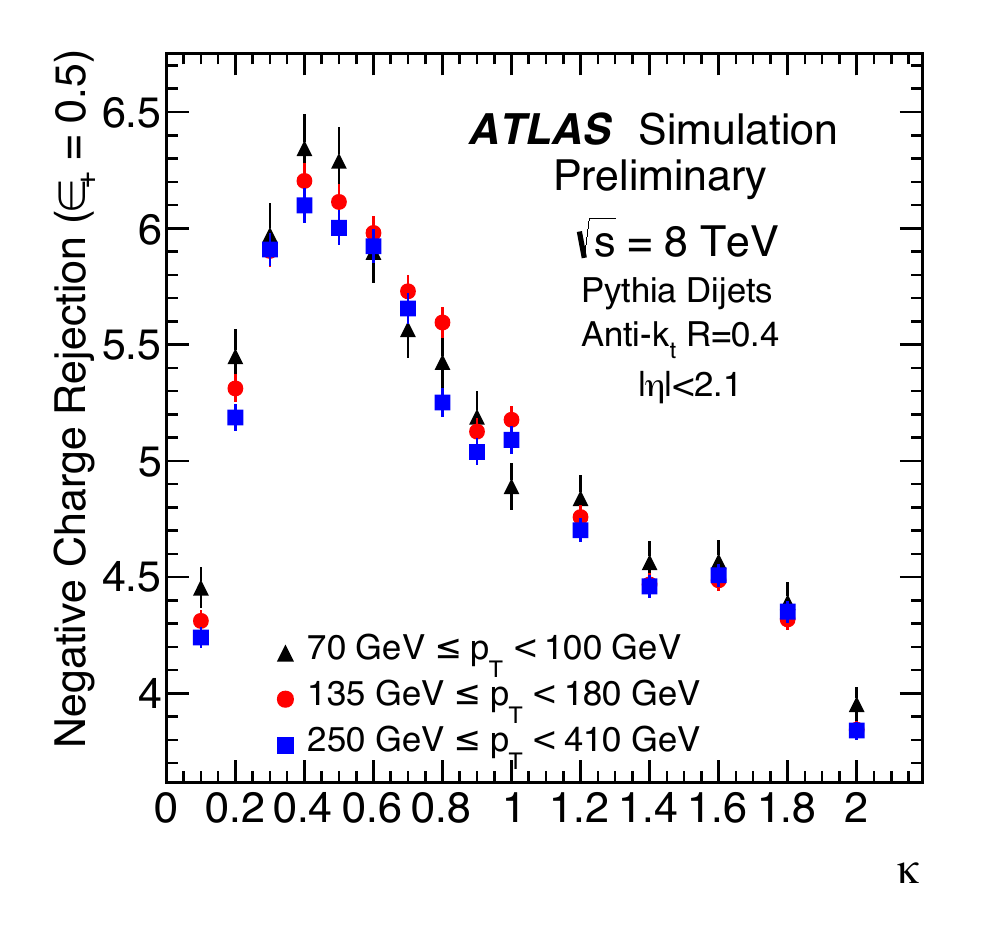}
\end{center}

\vspace{-7mm}

\caption{The $p_T$ dependance of the negative charge rejection at a fixed $\kappa=0.4$ for various positive charge efficiency working points (left) and the distribution of negative charge rejection as a function of $\kappa$ for a few $p_T$ bins (right).  The value $\kappa=0.4$ is optimal nearly independent of $p_T$.}
\end{figure}

\newpage

The sum of the jet charge of the leading two jets in QCD dijet events is shown in Fig. 6.  An equal composition of positive and negative charge flavor jets would have average zero, but the inclusive dijet spectrum is not an equal superposition of positive and negative charges.  As the momentum transfer increases, the valence quarks dominate.  Since the up valence quark has more probability than the down valence quark, the average jet charge increases with the dijet invariant mass (which is a proxy for $\sqrt{\hat{s}}$).  There are small systematic differences between data and MC, but further studies are required to understand this better.

\begin{figure}[h!]
\begin{center}
\includegraphics[width=0.5\textwidth]{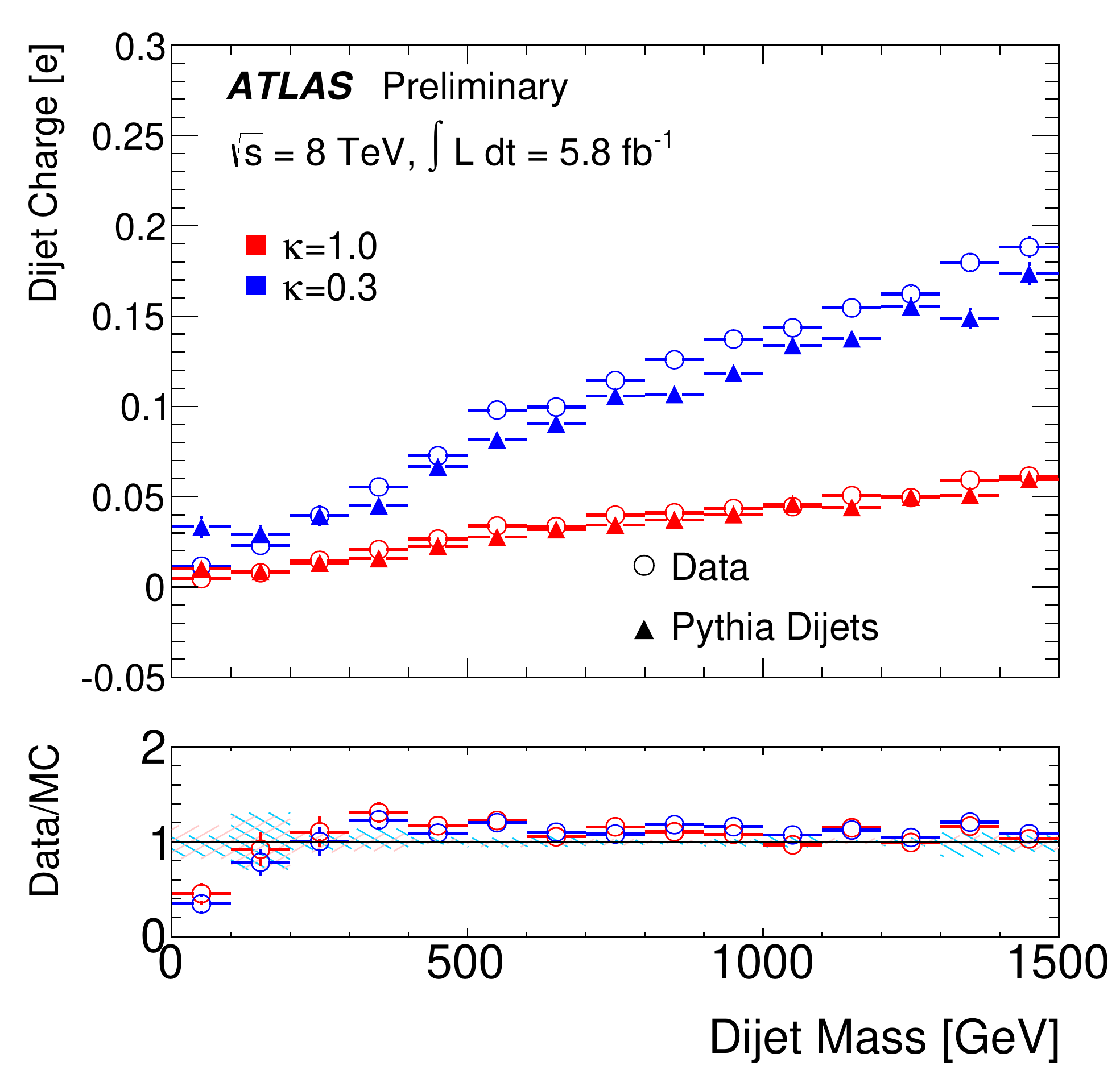}
\end{center}
\caption{The sum of the jet charges of the leading two jets in dijet events as a function of the dijet invariant mass for two values of the $p_T$ weighting factor $\kappa$.}
\end{figure}

\section{Conclusions}

The electric charge of quarks and gluons cannot be measured directly, but information about the charge is passed to the final state particles.  Jet charge is one observable sensitive to parton electric charge and has been shown in $t\bar{t}$ and dijet events to be well modeled by the MC.  In addition, a tag and probe study in $t\bar{t}$ events has shown directly that the jet charge can be used to tag the electric charge of a hadronically decaying $W$ boson.  In dijet events, the average jet charge increases with the energy scale of the collision, which is observed in both data and MC, though further studies are required to make a precise measurement and compare with calculations.  Jet charge has attractive properties and is complementary to other flavor-tagging variables -- many measurements and searches may benefit by adding it as a tool.

\Acknowledgements
This material is based upon work supported by the National Science Foundation Graduate Research Fellowship under Grant No. DGE-4747 and by a Stanford Graduate Fellowship.

\end{document}